\documentclass[%
 aip,
 amsmath,amssymb,
 reprint,%
]{revtex4-1}

\usepackage{graphicx}% Include figure files
\usepackage{dcolumn}% Align table columns on decimal point
\usepackage{bm}% bold math
\usepackage[utf8]{inputenc}
\usepackage[T1]{fontenc}
\usepackage{mathptmx}
\usepackage{etoolbox}

\makeatletter
\def\@email#1#2{%
 \endgroup
 \patchcmd{\titleblock@produce}
  {\frontmatter@RRAPformat}
  {\frontmatter@RRAPformat{\produce@RRAP{*#1\href{mailto:#2}{#2}}}\frontmatter@RRAPformat}
  {}{}
}%
\makeatother
\begin{document}

\preprint{AIP/123-QED}

\title[An Ultra-low-loss Compact Phase-Change Material-based Hybrid-mode Interferometer for Photonic Memories]{An Ultra-low-loss Compact Phase-Change Material-based Hybrid-mode Interferometer for Photonic Memories}
% Force line breaks with \\
\author{Ranjeet Dwivedi}
\email{ranjeetdwivedi2@gmail.com}
\affiliation{ 
Ecole Centrale de Lyon, CNRS, INSA Lyon, Université Claude Bernard Lyon 1, CPE Lyon, INL, UMR5270, 69130 Ecully, France%\\This line break forced with \textbackslash\textbackslash
}%
 
\author{Fabio Pavanello}% 

\affiliation{%
Univ. Grenoble Alpes, Univ. Savoie Mont Blanc, CNRS, Grenoble INP, CROMA, 38000 Grenoble, France %\\This line break forced% with \\
}%

\author{Regis Orobtchouk}
\affiliation{ 
Ecole Centrale de Lyon, CNRS, INSA Lyon, Université Claude Bernard Lyon 1, CPE Lyon, INL, UMR5270, 69130 Ecully, France%\\This line break forced with \textbackslash\textbackslash
}%

\date{\today}% It is always \today, today,
             %  but any date may be explicitly specified

\begin{abstract}
We propose a novel hybrid mode interferometer (HMI) leveraging the interference of hybridized TE-TM modes in a silicon-on-insulator (SOI) waveguide integrated with a GeSe phase change material (PCM) layer. The SOI waveguide's dimensions are optimized to support the hybridization of the fundamental transverse magnetic  ($TM_0$) and the first higher transverse electric ($TE_1$) mode. This design allows for efficient and nearly equal power coupling between these two modes, resulting in high-contrast interference when starting from the amorphous PCM state. The PCM's phase transition induces a differential change in the modal effective index, enabling high-contrast transmittance modulation. Our numerical simulations demonstrate a multilevel transmission with a high contrast of nearly 14 dB, when the amorphous region's length is varied incrementally, enabling multi-bit storage. The transmittance is maximized in the fully crystalline state with an insertion loss below 0.1 dB. The HMI can also operate as a quasi-pure phase shifter when partially amorphized, making it suitable for Mach-Zehnder interferometers. These characteristics make the proposed device a promising candidate for applications in photonic memories and neuromorphic computing. 
\end{abstract}

\maketitle

 Integrated photonic memories are emerging as a promising technology for energy-efficient and high-speed data storage, with potential applications as building blocks in neuromorphic computing \cite{li2019fast,carrillo2021system,carrillo2019behavioral,xia2024seven,wu2024freeform,tsakyridis2024photonic}. Memories based on a thin layer of phase-change material (PCM) placed above a silicon-on-insulator (SOI) waveguide, in contact or in close proximity, are regarded as one of the most promising approaches for next-generation photonic systems. Among the various optical PCMs, GeSbTe (GST) is the most widely studied due to its high refractive index contrast between amorphous and crystalline phases at 1550 nm wavelength. It enables compact device designs utilizing absorption modulation in various configurations\cite{li2019fast,carrillo2021system,carrillo2019behavioral, xia2024seven,yan2024high}. PCMs have been extensively utilized in Mach-Zehnder interferometers (MZIs), directional couplers, ring resonators, slot waveguides, etc., for, modulation, switching, mode conversion, and other reconfigurable photonic devices\cite{soref2015electro,dhingra2019design,chen2023non,chen2024deterministic,soref2024compact, zhang2020ultra,yan2024high,quan2022nonvolatile,song2022compact, wei2024inverse, zhang2023nonvolatile}. Specifically, MZIs require low-loss PCM-based phase shifters to achieve efficient phase modulation with minimal amplitude variation, translating into high extinction ratios. In particular, GST has been extensively explored utilizing incoherent architectures leveraging wavelength division multiplexing for neuromorphic applications \cite{carrillo2021system,xia2024seven,guo2022starlight}. However, its inherent high absorption limits its applicability in MZIs-based coherent photonic circuits, which require low-loss phase shifters \cite{tsakyridis2024photonic,harris2018linear}. Low-loss PCMs integrated on SOI waveguides can offer a nearly pure phase shift \cite{delaney2020new,fang2023arbitrary,teo2022comparison,zheng2020nonvolatile}, with minimal losses due to material absorption and mode field mismatch at the interface of the PCM patch. However, to obtain multilevel transmission, such structures need to be integrated into MZIs, directional couplers, or ring resonators, which increases their overall footprint.

In this paper, we propose a novel compact design based on a SOI waveguide integrated with GeSe, a low-loss PCM. This structure serves dual functions as both a quasi-pure phase modulator and an amplitude modulator, achieved by varying the crystallization fraction of the GeSe layer. In addition to offering a compact memory, it can serve as a phase shifter in neuromorphic coherent photonic circuits and as a variable optical attenuator in incoherent neuromorphic systems. Our approach leverages the interference between hybridized $TM_0 - TE_1$ modes in an SOI waveguide with optimized dimensions, enabling high-contrast amplitude modulation through the GeSe phase change. The proposed hybrid mode interferometer (HMI) demonstrates a high contrast of approximately 14 dB in its transmittance between crystalline and amorphous states, with an insertion loss (IL) below 0.1 dB when the PCM is in a fully crystalline state. A contrast exceeding 83\% is achieved across the wavelength range of 1.51 $\mu$m - 1.58 $\mu$m, with a maximum IL in the crystalline state of 0.46 dB. Our calculations indicate that the HMI can achieve multilevel transmittance, allowing for the storage of multiple bits of memory. Additionally, the performance of an MZI incorporating HMI in its arms is numerically evaluated, showing multilevel transmission and switching with an IL of approximately 0.7 dB. The results are also compared with those of an HMI and MZI using GST as a PCM. The proposed cross-section can be readily adapted in a CMOS-compatible silicon photonics platform \cite{10173974}.

\begin{figure}[h!]
    \centering
    \includegraphics[width=\linewidth]{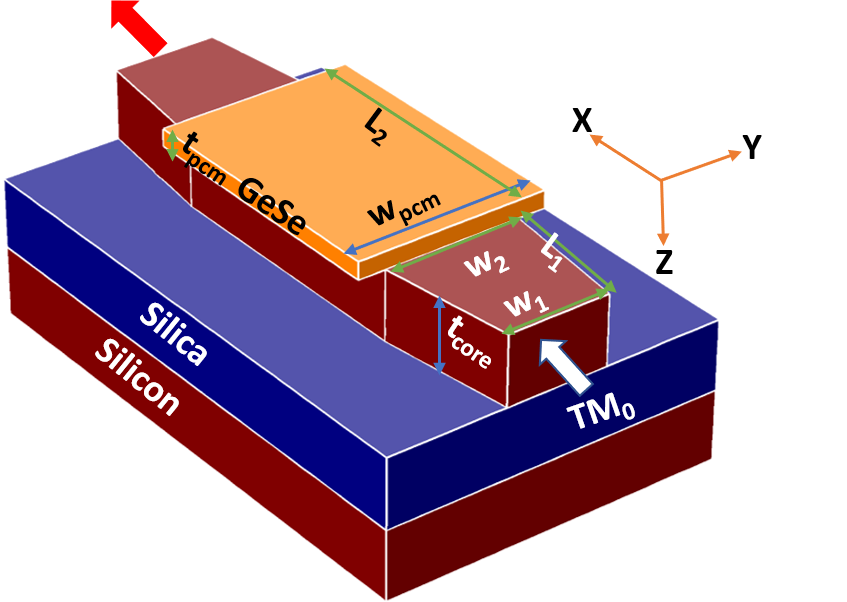} 
    \caption{Schematic of the proposed hybrid mode interferometer.}
    \label{fig:Fig1}
\end{figure}
Figure \ref{fig:Fig1} illustrates the schematic of the considered hybrid mode interferometric structure, which consists of a waveguide with 50-nm-thick GeSe PCM integrated on top of a standard strip SOI  interconnected between identical tapered input/output SOI waveguides. The silicon and buried oxide (BOX) layers have thicknesses of 0.22 µm and 2.0 µm, respectively. The structure is considered to be surrounded by silica in the cladding region. 
The width of the tapered section is considered to vary between  \( w_1 \) and \( w_2 \) across its length \( L_1 \). The middle section has a uniform width \( w_2 \) across its length \( L_2 \). The width of the PCM layer is kept fixed at 1.0 µm. The core dimensions of the input and output ends are selected to support a single TE/TM mode. The dimensions of the middle waveguide are optimized to support hybridized TE-TM modes. We employed the finite difference method using Lumerical Mode solver to calculate the effective indices and polarization fractions. The refractive indices of silicon and silica are taken from Palik's data \cite{palik1998handbook}. The refractive index of the PCM layer in its amorphous and crystalline states are taken as 2.4 + 0.00006i and  2.97 + 0.00006i \cite{soref2015electro}.\\

\begin{figure}[h!]
    \centering
    \includegraphics[width=\linewidth, trim=1.5cm 0cm 2cm 0cm,clip]{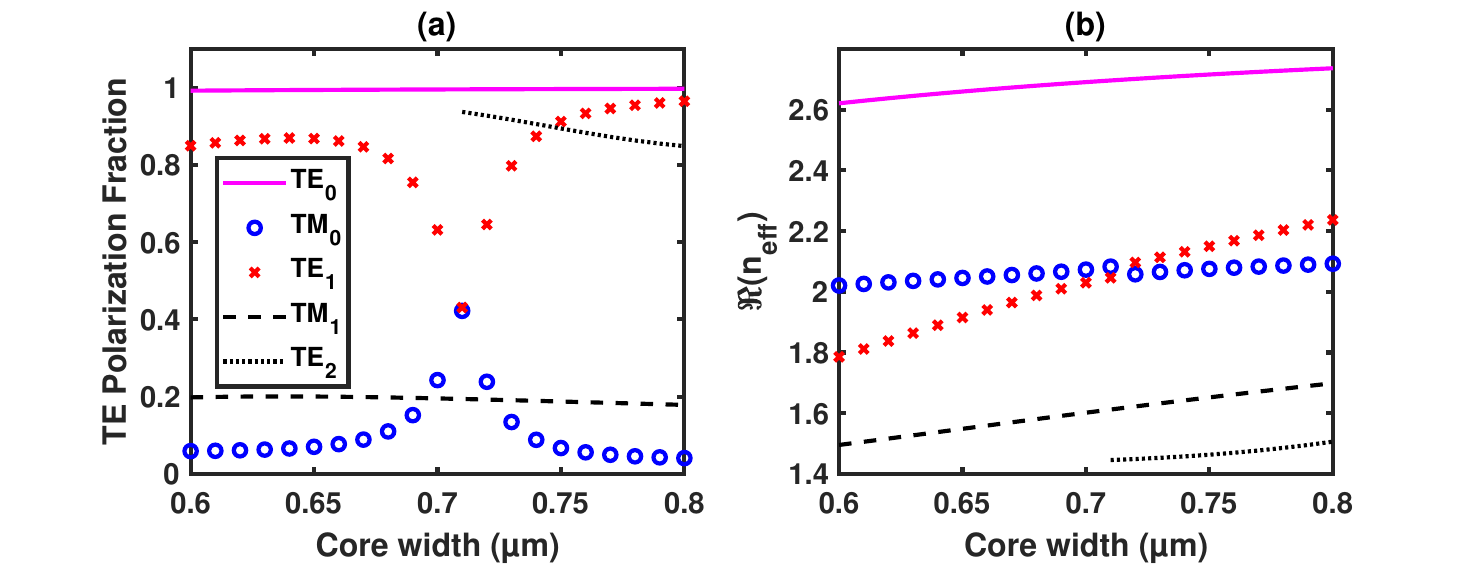} 
    \caption{Variation of (a) TE polarization fraction and (b) real part of the effective index of the different guided modes with the core width, for the middle waveguide section. Legend in (a) also applies to (b). }
    \label{fig:Fig2}
\end{figure}
 Figure \ref{fig:Fig2}(a) shows the variation of the transverse electric  (TE) polarization fraction among the various guided \(TE\) and \(TM\) modes. It should be noted that the TE polarization fraction refers to the proportion of modal power carried by the y-component of the electric field, as the waveguide's cross-section lies in the y-z plane. In our calculations, the wavelength (\( \lambda \)) is 1.55 µm, and \( w_2 \) is varied in the range of 0.6 µm – 0.8 µm. The PCM layer is considered to be initially in the amorphous state. It can be seen that the TE polarization fraction of the \(TM_0\)  (\(TE_1\)) mode first increases (decreases), reaching a maximum (minimum) at \( w_2 = 0.71 \) µm, and then decreases (increases) after that. At \( w_2 = 0.71 \) µm, the \(TM_0\) and \(TE_1\) modes are hybridized and have nearly equal TE polarization fractions of 42\%. The real part of the effective indices of these two modes becomes very close in the hybridization region, as evident in  Fig. \ref{fig:Fig2}(b). The TE polarization fraction of the \(TE_0\) and \(TE_2\) modes, respectively, remain above 99\%  and 84\% while that for the \(TM_1\) varies between 17\% to 20\%.

\begin{figure}[h!]
    \centering
    \includegraphics[width=\linewidth, trim=0.8cm 0cm 1.3cm 0.4cm,clip]{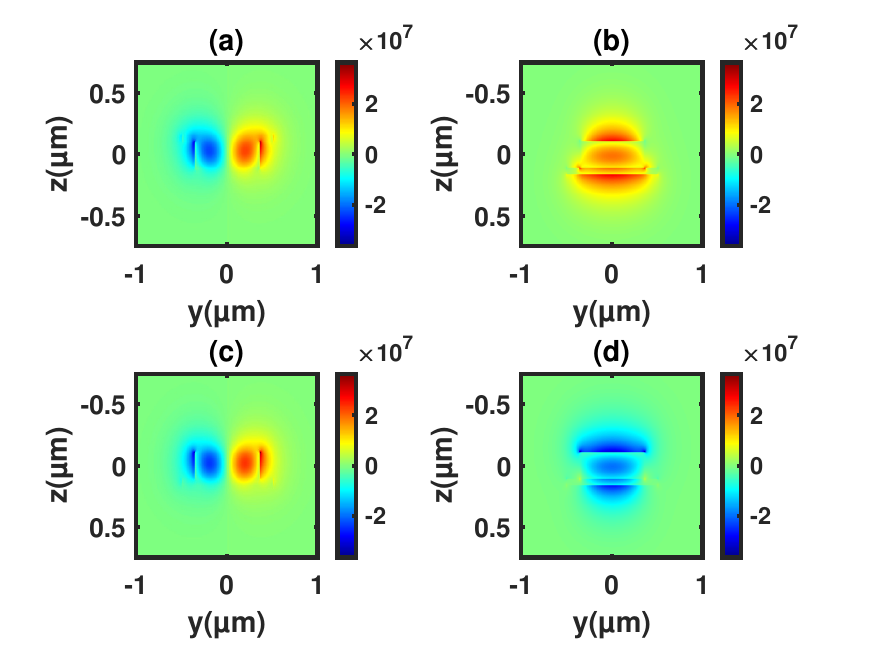} 
    \caption{Spatial distribution of the $E_y$ and $E_z$ electric field components, respectively, for $TM_0-TE_1$ hybrid mode 1 (a, b) and mode 2 (c, d).}
    \label{fig:Fig3}
\end{figure}

At \( w_2 = 0.71 \) µm, the y and z components of electric fields (denoted by $E_y$ and $E_z$) of the \(TM_0\)\(-TE_1\) hybrid modes, are shown in Fig. \ref{fig:Fig3}. This demonstrates that the $E_y$/$E_z$ of both modes displays nearly similar patterns while the $E_z$ of one mode, when inverted with respect to the \textit{z}-axis, resembles $-E_z$ of the other hybrid mode.  

To calculate the transmittance of the structure, a TM-polarized input light is considered to excite the fundamental TM mode in the input taper. After propagating to the central waveguide section, light is coupled to the hybridized TE-TM modes. A fraction of the power coupled to the hybridized modes is then back-coupled to the fundamental TM mode at the output end of the last taper, contributing to the output power.

In our calculations, we have taken \( w_1 = 0.4 \) µm, \( w_2 = 0.71 \) µm, and \( L_1 = 10 \) µm for the tapered section to ensure efficient power coupling between the hybridized modes. Mode overlap calculations show that the power is efficiently coupled to the forward-propagating hybridized modes, leading to negligible reflection at the interfaces. In such a case, the transmittance can be approximated with the interference between these two modes. Therefore, the transmittance can be written as:
\begin{equation}
T \approx \left| C_{1} e^{i \beta_{1} L_2} + C_{2} e^{i \beta_{2} L_2} \right|^2
\label{eq:1}
\end{equation}
Here, \(\left|C_{1}\right|\) (\(\left|C_{2}\right|\)) represents the magnitude of the product of the modal overlaps between the hybridized $TM_0-TE_1$ modes of the central section of the HMI with the fundamental TM mode of the taper at the input and output interfaces. This corresponds to the fraction of power coupled to (from) the $TM_0-TE_1$ hybrid modes having propagation constants $\beta_{1}$ and $  \beta_{2}$ from (to) the input (output) taper. The transmittance calculated from Eq. (\ref{eq:1}) closely matches the results from the eigenmode expansion method, with an error ranging between 0.05\% and 2.8\% across the wavelength range 1.45 µm - 1.65 µm.

\begin{figure}[h!]
    \centering
    \includegraphics[width=\linewidth, trim=2.2cm 0cm 2.2cm 0cm,clip]{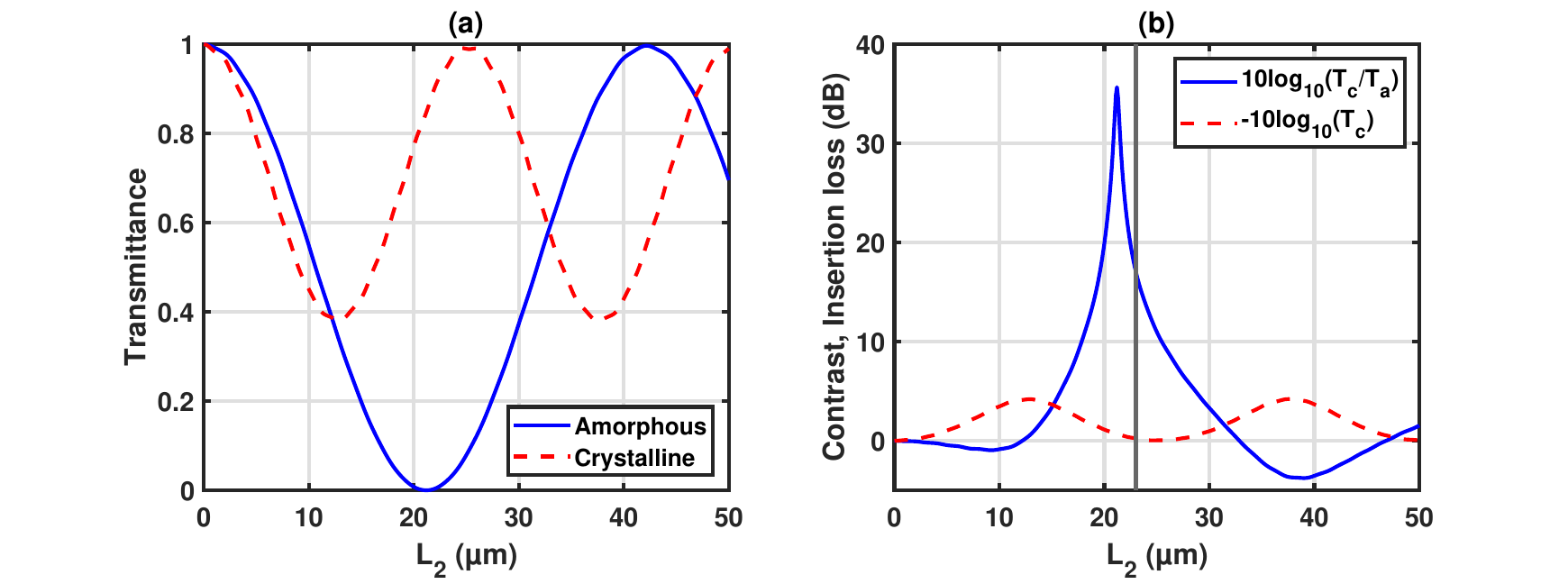} 
    \caption{Variation of (a) transmittance in amorphous (solid blue) and crystalline (dashed red) states, and (b) contrast (solid blue) and insertion loss (dashed red) with the HMI length.}
    \label{fig:Fig4}
\end{figure}

The transmittance calculated using the eigenmode expansion method-based Lumerical  EME solver is shown in Fig. \ref{fig:Fig4}(a) for both amorphous and crystalline phases. It can be seen that the transmittance has nearly periodic variation with \( L_2 \) in both states, which is a result of the interference between the hybrid modes. The transmission contrast in the amorphous case is $\approx$ 1 due to nearly equal power coupling to the hybrid modes from the tapered section. In the crystalline phase, the polarization fraction of these modes is different, resulting in unequal power coupling and a decrease in contrast. Moreover, the period is different in both states due to different changes in the effective indices of the two modes, leading to a change in the phase difference. Figure \ref{fig:Fig4}(b) shows the variation of transmission contrast between the crystalline and amorphous states and the IL in the crystalline state. It is found that the contrast is maximum at \( L_2 = 21.2 \) µm, where the transmittance in the amorphous state ($T_a$) is minimum. The IL, however, is minimum at \( L_2 = 24.6 \) µm, where the transmittance in the crystalline state ($T_c$) is maximum. For \( L_2 \) varying between 21.2 µm and 24.6 µm, the contrast ranges from 35.6 dB to 12 dB, while the IL varies between 0.7 dB and 0.04 dB. Therefore, depending on the requirement, we can select either very high contrast with increased IL or somewhat lower contrast with decreased IL. Given this, we take \( L_2 = 23.9 \) µm, where the contrast is $\approx$ 14 dB and the IL is below 0.1 dB.  

\begin{figure}[h!]
    \centering
    \includegraphics[width=\linewidth, trim=1.9cm 0cm 2.0cm 0cm,clip]{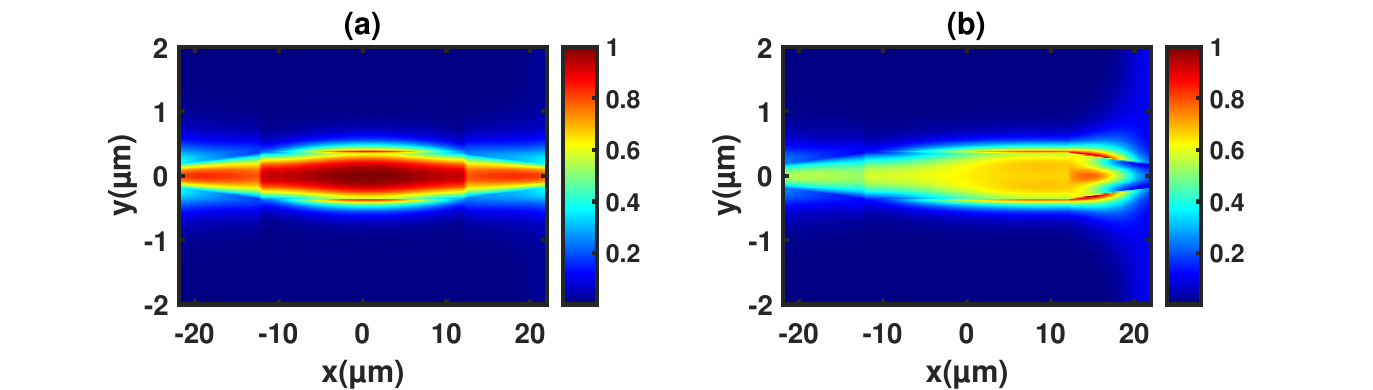} 
    \caption{Spatial map of the electric field magnitude in the x-y plane passing through the center of the waveguide's core in (a) crystalline and (b) amorphous states, respectively.}
    \label{fig:Fig5}
\end{figure}

The electric field distribution in the x-y plane passing through the center of the core is shown in Fig. \ref{fig:Fig5}, clearly representing the high and low transmittance in the crystalline and amorphous states, respectively.

\begin{figure}[h!]
    \centering
    \includegraphics[width=\linewidth]{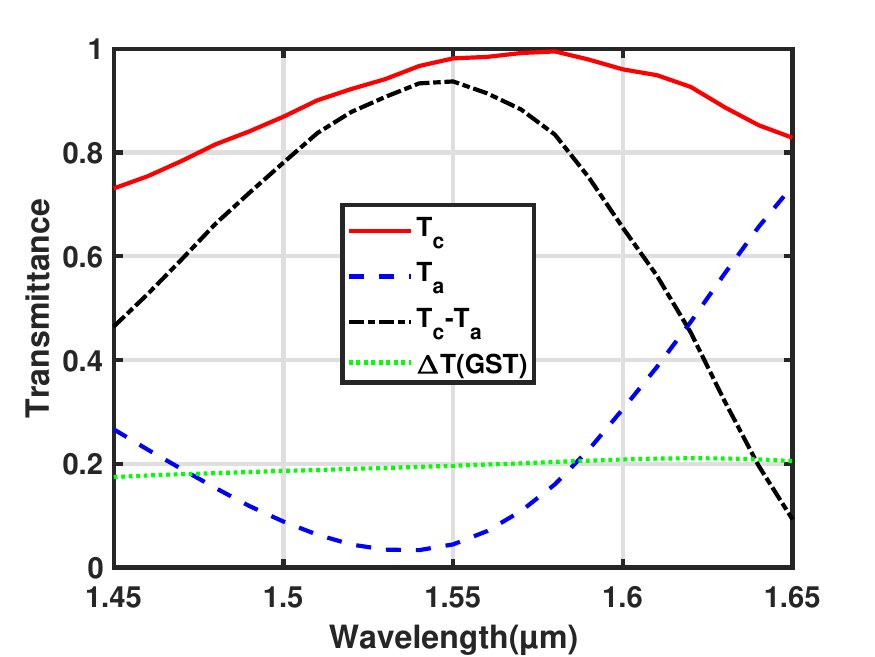}
     \caption{Spectral variation of transmittance of HMI in crystalline (solid red) and amorphous (dashed blue) state of GeSe. The transmission contrast is also shown on the same graph for the GeSe (dash-dotted black) and GST (dotted green).}
    \label{fig:Fig6}
\end{figure}

Next, we calculated the spectral dependence of the proposed HMI. Figure \ref{fig:Fig6} presents the transmission spectra of the HMI  in both crystalline and amorphous phases.  The transmission contrast ($T_c-T_a$) is also plotted in Fig. \ref{fig:Fig6}. The variation in transmittance is governed by the spectral dependence of the amplitudes and the phase difference between the two hybridized modes.
In the crystalline phase, the HMI exhibits a transmission peak around \( \lambda = 1.58 \) µm, where the condition for constructive interference between the two hybrid modes is closely achieved. The decrease in transmittance by moving away from \( \lambda = 1.58 \) µm corresponds to deviations from the constructive interference condition.
In the amorphous state, a transmission minimum equal to 0.033 is observed at \( \lambda = 1.54 \) µm, where the conditions for destructive interference and equal power coupling between the two modes are closely satisfied. The increase in transmittance in this case is associated with deviations from the destructive interference condition and increasing difference in power coupling between the hybridized modes from the input taper. The maximum contrast is found to be 0.937 with an IL of 0.09 dB (for crystalline GeSe) at \( \lambda = 1.55 \) µm. In the wavelength range of 1.51 µm to 1.58 µm, the HMI demonstrates a contrast of over 83$\%$ and a worst-case IL of 0.46 dB. 
We also present a comparison of the transmittance contrast of an HMI incorporating GST as the PCM. The refractive index data of GST in the crystalline and amorphous states are taken as 6.11 + 0.83i  and 3.94 + 0.045i \cite{yan2024high}.  In this case, the thickness of the GST layer is fixed at 20 nm, resulting in an optimum width \( w_2 = 0.69 \) µm that supports mode hybridization in the amorphous state. The length \( L_2 \) is taken to be 40 µm to achieve a maximum near the constructive interference regime at \( \lambda = 1.55 \) µm. Notably, due to the significant propagation loss related to the much higher losses in GST, the maximum is not solely dependent on the accumulated phase difference but also on the amplitudes of the modes. For the considered length, the value of transmittance in the crystalline state of GST is negligible, leading to a transmission contrast (\( \Delta T \)) equal to the transmittance in the amorphous state. It should be noted that \( \Delta T \) varies between 0.175 and 0.205, which is lower than the contrast in the case of HMI with GeSe within the wavelength range 1.45 µm - 1.63 µm. The increase in contrast on the higher wavelength side is attributed to the decrease in modal propagation loss. Moreover, in the 1.51 µm to 1.58 µm wavelength range, the contrast in the case of GST is significantly less compared to that of GeSe.

\begin{figure}[h!]
    \centering
    \includegraphics[width=\linewidth, trim=1cm 0cm 1.5cm 0cm,clip]{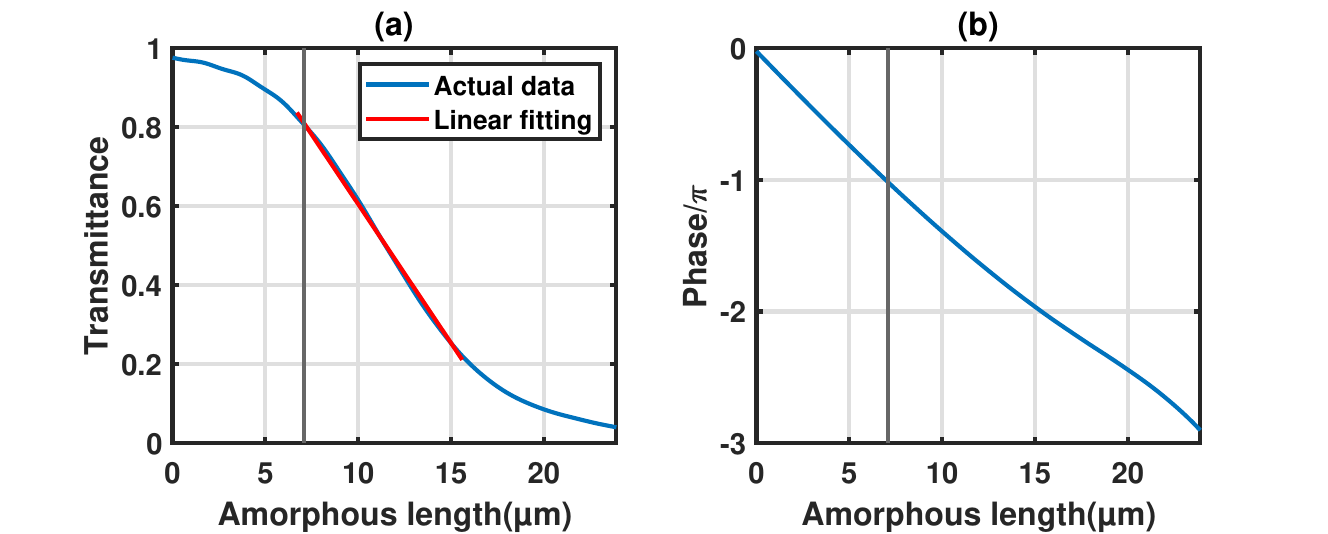} 
   \caption{Variation of (a) transmittance and (b) phase of the transmission coefficient with the length of the amorphous region.}
    \label{fig:Fig7}
\end{figure}

After establishing a high-contrast optical transmission between the crystalline and amorphous states of the GeSe PCM, we next explore the potential of the HMI for multilevel photonic memory. For this purpose, we considered the PCM layer to be divided into multiple identical cells, which can change their phase from crystalline to amorphous in an incremental manner. In our calculations, we have considered the PCM layer to be divided into 239 cells each having 100 nm length. The transmittance in this case is calculated using Lumerical EME solver and is shown in Fig. \ref{fig:Fig7}(a). The PCM layer is initially considered to be in a fully crystalline state, leading to a maximum transmittance of 0.979 (-0.09 dB). The gradual switching, e.g., by using local microheaters, of the PCM cells from their initial crystalline state increases the amorphous region across the length of the PCM layer, resulting in a decrease in transmittance. The transmittance scales almost linearly in the range 83.5 \% to 21 \%  for the amorphous length in the range 6.7 µm - 15.6 µm (marked by solid red line obtained from linear fitting of data in the range 6.7 µm - 15.6 µm). A transmittance minimum of 0.04 (-14 dB) is achieved when the PCM is fully amorphous leading to a contrast of \(\approx\) 14 dB.
Figure \ref{fig:Fig7}(b) shows the variation of the phase of the transmission coefficient with the length of the amorphous region. Interestingly, an amorphized region of only 7.1 µm leads to nearly $\pi$ phase difference compared to the fully crystalline state. This indicates that an HMI with partial amorphization could be incorporated in the arms of an MZI.
 
\begin{figure}[h!]
    \centering
    \includegraphics[width=\linewidth]{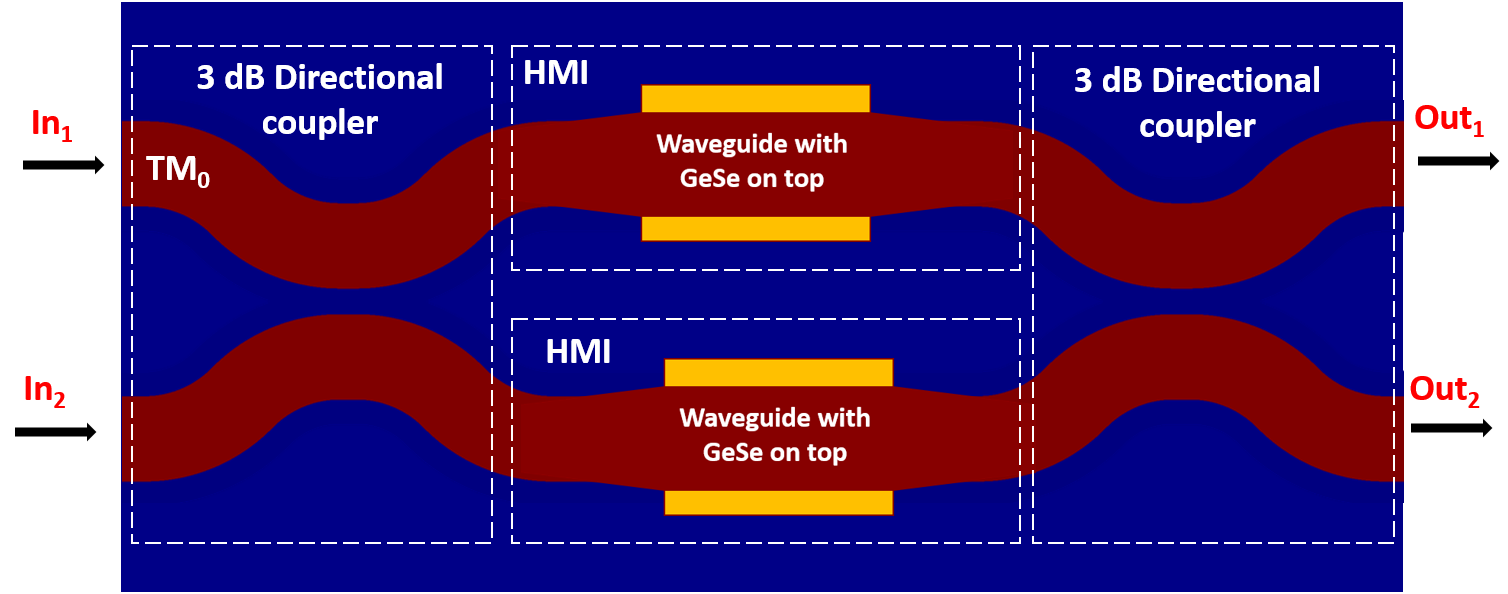} 
    \caption{2x2 MZI configuration with 3 dB directional couplers and HMI in each arm.}
    \label{fig:MZI}
\end{figure}

\begin{figure}[h!]
    \centering
    \includegraphics[width=\linewidth, trim=1.5cm 0cm 2cm 0cm,clip]{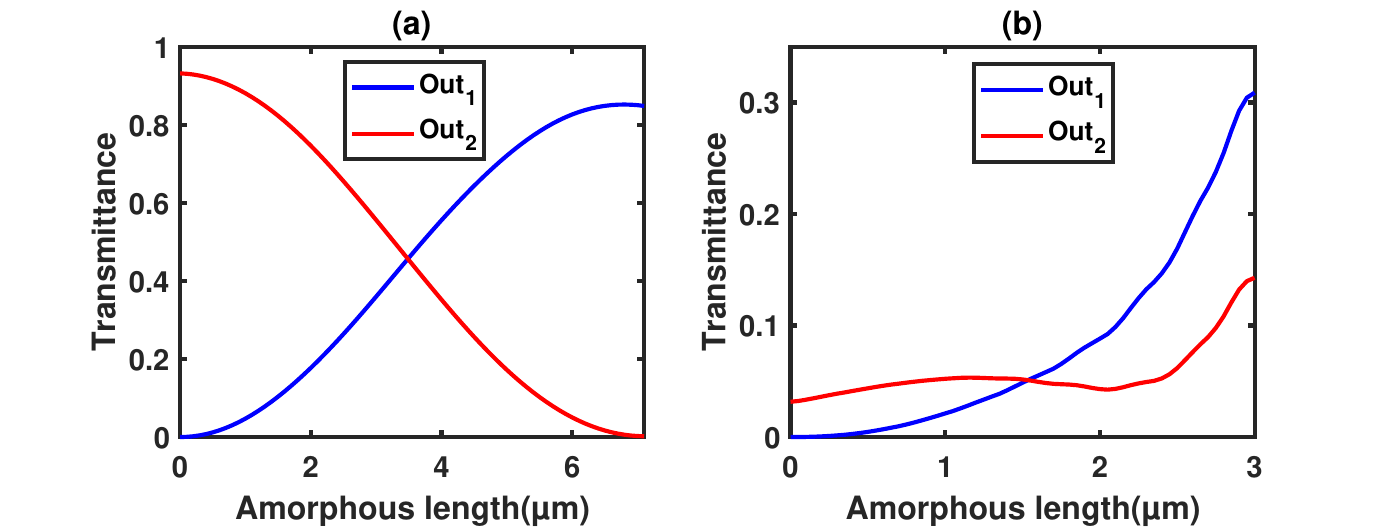} 
    \caption{Variation in transmittance of an MZI with (a) HMI and (b) GST-based arms, as a function of increasing amorphous region in one of the arms.}
    \label{fig:Fig9}
\end{figure}

Finally, we explored the transmission characteristics of an MZI using HMIs in its arms, as shown in Fig. \ref{fig:MZI}. Light (TM polarized) is considered to be launched only in the first input port exciting the $TM_0$ mode. In our calculations, the IL of the couplers is taken as 0.1 dB. One of the MZI arms is fixed in the crystalline state, while the length of the amorphous region is varied in the other arm. 
The transmittance at both output ports of the MZI is shown in Fig. \ref{fig:Fig9}(a). It can be seen that, with the change in the amorphous/crystalline fraction, the transmittance can be modulated. As expected, a 7.1 µm long amorphous region leads to a power switching between the two output ports due to the $\pi$ phase difference between the two arms. The highest IL is found to be $\sim$ 0.7 dB. For comparison, the transmittance of an MZI incorporating GST in its arms is shown in Fig.  \ref{fig:Fig9} (b). The injection mode at the first input port of the MZI is again a $TM_0$ mode. However, unlike the GeSe-based MZI, we did not utilize a GST-based HMI structure in the MZI arms. Instead, we employed a SOI waveguide integrated with a 20 nm thick, 3 µm long GST patch to achieve lower IL. The width of the SOI waveguide and that of GST is taken as 400 nm. For the considered dimensions of the GST, nearly $\pi$ phase difference is obtained between the transmission coefficients of the crystalline and amorphous states. It can be seen that the transmission at both ports is very low when both arms are in a fully crystalline state. With the increase in the length of the amorphous region, the total transmittance is increased reaching a maximum of around 45$\%$ when the GST layer in one of the arms becomes fully amorphous. The IL in this case is found to vary between 15 dB and 3.5 dB due to the imbalance created by the high absorption loss of GST.

Therefore, it is clear that using GST in an MZI configuration suffers from high IL and low contrast compared to GeSe-based approaches.
The IL of a GeSe-based MZI can be further minimized by employing a simple GeSe-based phase shifter, without an HMI. For instance, a GeSe patch with a thickness of 50 nm and a width of 400 nm integrated on an SOI waveguide with the same width and a thickness of 220 nm can achieve a $\pi$ phase shift between the crystalline and amorphous states over a patch length of 8 µm. When this phase shifter is incorporated into the MZI arms, it enables power switching between the output ports, with an IL of 0.54 dB compared to 0.7 dB in the HMI-based MZI configuration. It is important to note that the IL in the HMI-based MZI arms is primarily due to slight amplitude variations in the transmission during phase shifts. On the other hand, the IL in simple GeSe patch-based MZI arms is mainly caused by mode field mismatches at the GeSe interfaces. The proposed HMI device has a compact footprint of approximately 44 $\mu m^2$, making it considerably smaller than a standard 1×1 MZI, which incorporates 1×2 power splitters at both the input and output, each occupying over 50 $\mu m^2$ space \cite{9476033}. These memories can be written either optically, using in-plane or out-of-plane optical pulses, or electrically, via integrated heaters.

We proposed the design of a novel hybrid mode interferometer exploiting the interference between hybridized $TM_0 - TE_1$ modes of an SOI waveguide integrated with GeSe PCM. For the optimized dimensions, the HMI shows a high transmission contrast between the crystalline and amorphous states of GeSe. Moreover, a contrast above 83\% is obtained in the wavelength range 1.51 µm - 1.58 µm. The multilevel transmission performance is simulated considering a gradual increase in the length of the amorphous region showing a contrast of $\approx$14 dB between fully crystalline and fully amorphous GeSe, with a fully crystalline state IL < 0.1 dB at $\lambda=1.55$ µm. The HMI can act as a quasi-pure phase shifter with partial amorphization enabling its integration into MZIs.  These features make the HMI device a promising candidate for multi-bit photonic memories, variable optical attenuators, and phase shifters for energy-efficient neuromorphic photonic computing. The proposed device has a quite small footprint (43.9 µm $\times$ 1 µm ) compared to a standard 1×1 MZI.

\section*{Acknowledgments}
This work has received funding from the European Union’s Horizon
Europe research and innovation program under the grant agreement
No. 101070238.
\section*{AUTHOR DECLARATIONS}
\subsection*{Conflict of Interest}
The authors have no conflicts to disclose.
\subsection*{Author Contributions}
\textbf{Ranjeet Dwivedi:} Conceptualization (equal); Data curation (lead); Formal analysis (equal); Methodology (lead); Software (lead); Visualization (lead); Writing – original draft (lead); Writing – review \& editing (equal). \textbf{Fabio Pavanello:} Conceptualization (equal); Formal analysis (equal); Funding acquisition (equal); Methodology (supporting); Project Administration (lead); Supervision (equal); Writing – review \& editing (equal). \textbf{Regis Orobtchouk:} Conceptualization (equal); Formal analysis (equal); Funding acquisition (equal); Methodology (supporting); Supervision (equal); Writing – review \& editing (equal).

\section*{DATA AVAILABILITY STATEMENT}
The data that support the findings of this study are available
from the corresponding author upon reasonable request.
\section*{References}
\bibliography{paper}
\end{document}